\documentclass[aps,prl,twocolumn,showpacs]{revtex4}
\usepackage{graphicx}
\usepackage{wasysym}

\begin{document}
\def\be{\begin{equation}}
\def\ee{\end{equation}}

\title{Activity-dependent brain model explaining EEG spectra}

\author{Lucilla de Arcangelis,$^1$ Hans J. Herrmann$^2$ and
Carla Perrone-Capano$^3$}
\affiliation{$^1$
Department of Information Engineering and INFM-Coherentia,
Second University of Naples, 81031 Aversa (CE), Italy  \\
$^2$ Institute for Computer Applications 1, University of Stuttgart, 
Pfaffenwaldring 27, D-70569 Stuttgart, Germany \\
$^3$ University of Catanzaro "Magna Graecia", Dept. of Pharmacobiology, 
I-88021 Roccelletta di Borgia (CZ), and IGB "A. Buzzati Traverso", CNR, 
Via P. Castellino 111, I-80131 Napoli, Italy\\}

\begin{abstract}
Most brain models focus on associative memory or calculation capability, 
experimentally inaccessible using physiological methods. Here we present a 
model explaining a basic feature of electroencephalograms (EEG). Our model is 
based on an electrical network with threshold firing and plasticity of 
synapses that reproduces very robustly the measured exponent 0.8 of the 
medical EEG spectra, a solid evidence for self-organized criticality. 
Our result are also valid on small-world lattices. 
We propose that an universal scaling behaviour 
characterizes many physiological signal spectra for brain controlled 
activities.
\end{abstract}


\maketitle

\vskip 1cm

One of the most astonishing properties of the brain is its plasticity, i.e.
the ability to modify the structural and functional properties of synapses,
occurring mostly during development and learning \cite{alb}.
The mammalian central
nervous system relies on precise synaptic circuits to function correctly.
These circuits are assembled during development by the formation of synaptic
connections between hundreds of thousands of neurons. Although molecular
interactions direct the early formation of circuitry, this initial patterning
is followed by a prolonged period during which the establishment of highly
organized synaptic circuits in the developing human brain is thought to depend
on neural activity. This transforms immature circuits into the organized
connections that subserve adult brain function \cite{kat}.
In the central nervous
system much of this plastic sculpting of neuronal connections is thought
to occur during "critical periods" of early postnatal life \cite{hen},
when circuits
are particularly susceptible to electrical activity triggered by external
sensory inputs \cite{kat2}.

The most compelling and reliable models of activity-dependent synaptic
plasticity in the brain are Long Term Potentiation (LTP) and Long Term
Depression (LTD): persistent increases and decreases
in synaptic efficacy that can be elicited in mammalian neurons, based on
recent patterns of activity. LTP of synaptic transmission is traditionally
elicited by synchronous, high-frequency inputs, whereas LTD typically occurs
following repeated low frequency afferent stimulation \cite{pau}.
Strong evidence
suggests that LTP and LTD are candidate mechanisms mediating activity-dependent
synaptic plasticity during brain development \cite{kat},
and many forms of adaptive
behaviour, including learning and memory \cite{bra}.

In the last years many different time series emerging from neural activities 
\cite{alb} have been analysed through power spectra and generically 
power-law decay 
has been observed. This behaviour remains unexplained. Understanding its 
origin is not only a major theoretical challenge but also of eminent 
importance in many applications, in particular to give a solid basis to the 
interpretation of EEG \cite{gev,buz}. 
A large number of time series analyses have been performed 
on medical data that are directly or indirectly related to brain activity. 
Prominent examples are EEG data which are used by neurologists to discern 
sleep phases, diagnose epilepsy and other seizure disorders as well as brain 
damage and disease \cite{gev}. An other  example of a physiological 
function which 
can be monitored by time series analysis is the human gait which is controlled 
by the brain \cite{hau}. For all these time series the power spectrum, i.e. 
the 
square of the amplitude of the Fourier transformation double logarithmically 
plotted against frequency, generally features a power law at least over one 
or two orders of magnitude with exponents between 1 and 0.7. On top of this 
background power law, additional structures give information on the details 
of the pathology and can point to specific resonance, frequency cut-offs and 
other deviations. While much focus is given to these secondary structures, 
the basic power law remains largely unexplained.

Models for brain activity based on the convolution of oscillators \cite{ash} or 
stochastic waiting times \cite{iva} have been proposed. They are 
essentially abstract 
representations on a mesoscopic scale, but none of them is based on the 
behaviour of a neural network itself. In order to get real insights on the 
relation between these time series and the microscopic, i.e. cellular, 
interactions inside a neural network, it is necessary to identify the 
essential ingredients of the brain activity responsible for characteristic 
scale-free behaviour observed through the power law of the spectrum, as 
discussed above. This insight is the basis for any further understanding of 
the diverse additional features that are observed and interpreted by 
practitioners that analyse these time series for diagnosis. Therefore the 
formulation of the right brain model that yields the correct power spectrum 
is of crucial importance for any further progress in the understanding of 
the living brain.

Here we report on a new model that captures the three most important 
ingredients yielding the expected power law, namely threshold firing, synapse 
adaption and network plasticity, including both LTP and LTD. 
Despite its simplicity our model reproduces 
with astonishing precision the experimentally observed exponent of the power 
spectrum, already for rather small networks. In agreement with real data, 
this exponent turns out to be extremely robust against modifications of the 
various parameters of the model. With this result we claim having made a 
breakthrough in the generic understanding of the diverse electrical time series.

We consider a simple square lattice of size $L\times L$ on which each site 
represents the cell body of a neuron, each bond a synapse. Therefore, on 
each site we have a potential $v_i$ and on each bond a conductance $g_{ij}$. 
Whenever at time $t$ the 
value of the potential at a site $i$ is above a certain threshold 
$v_i \geq v_{\rm max}$, 
approximately equal to $-55mV$ for the real brain, the neuron fires, i.e. 
generates an "action potential", distributing charges to its connected 
neighbours in proportion to the current flowing through each bond
$$v_j(t+1)=v_j(t)+ v_i(t) {i_{ij}(t)\over\sum_k i_{ik}(t)} \eqno (1)$$
where $v_j(t)$ is the potential at time $t$ of site $j$, nearest neighbor of 
site $i$, $i_{ij}= g_{ij} (v_i-v_j)$ and the sum is extended to all nearest 
neighbors $k$ of site $i$
that are at a potential $v_k < v_i$. 
The conductances are initially all set equal to unity whereas the neuron 
potentials are uniformly distributed random numbers between 
$v_{\rm max} - 2$ and $v_{\rm max} - 1$. 
The potential is fixed to zero at top and bottom whereas periodic
boundaries are imposed in the other direction.

The system is stimulated at one input (source), a site in the centre of the 
lattice, and the electrical activity is monitored as function of time by 
measuring the total current flowing in the system. The firing rate of real 
neurons is limited by the refractory period, i.e. the brief period after the 
generation of an action potential during which a second action potential is 
difficult or impossible to elicit. The practical implication of refractory 
periods is that the action potential does not propagate back toward the 
initiation point and therefore is not allowed to reverberate between the cell 
body and the synapse. In our model, once a neuron fires, it remains quiescent 
for one time step and it is therefore unable to accept charge from firing 
neighbours. This ingredient indeed turns out to be crucial for a controlled 
functioning of our numerical model. In this way an avalanche of charges can 
propagate far from the input through the system similarly to the dynamics of 
self-organized critical systems \cite{bak}, 
as observed in organotypic cultures from 
coronal slices of rat cortex \cite{beg} where neuronal avalanches are 
stable for many hours \cite{beg2}.

\begin{figure}
\includegraphics[width=6cm,angle=270]{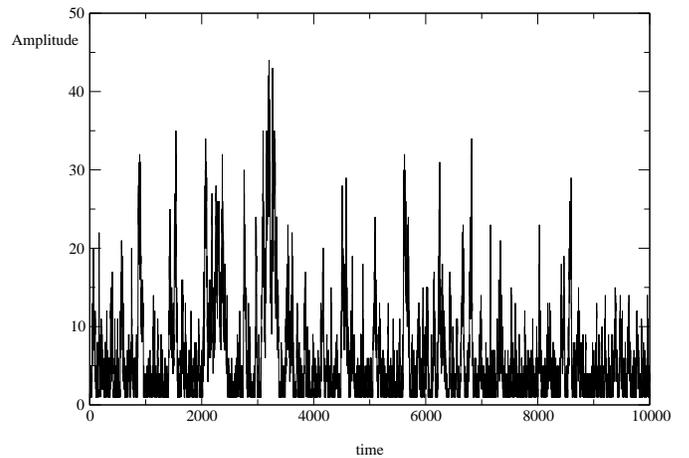}
\caption{ Total current flowing in one configuration of lattice ($L = 1000$)
as function of time in a sequence of several thousand stimuli. The value of
the parameters is $\alpha = 0.03$, $\sigma_t = 0.0001$, $v_{\rm max} = 6$.
\\}
\label{fig1}
\end{figure}

Every site that at a given time $t$ is at or above threshold $v_{\rm max}$ 
fires according to eq. (1), then the conductance of all the bonds that have 
carried a current is increased in the following way
$$g_{ij}(t+1) =g_{ij}(t) +\delta g_{ij} (t) \eqno (2)$$
where $\delta g_{ij}(t)=k \alpha  i_{ij}(t)$, with $\alpha$ being a 
dimensionless parameter and $k$ a unit constant bearing the dimension 
of an inverse potential. After applying eq. (2) the time variable of our 
simulation is increased by one unit. 
Eq. (2) describes the LTP mechanism, whose strength
is tuned by the parameter $\alpha$. 
Once an avalanche of firings comes to an end, the 
conductance of all the bonds with non-zero conductance is 
reduced by the average conductance increase per bond, 
$$\Delta g = \sum_{ij, t} \delta g_{ij} (t)/ N_b \eqno (3)$$
where $N_b$ is the number of bonds with non-zero conductance. Eq. (3)
implements the LTD process. The quantity $\Delta g$
depends on $\alpha$ and on the response of the brain to a given 
stimulus. In this way our electrical network "memorizes" the most used paths 
of discharge by increasing their conductance, whereas the less used synapses 
atrophy. Once the conductance of a bond is below an assigned small value 
$\sigma_t$, we remove it, i.e. set it equal to zero, which corresponds to 
what is known as pruning. This remodelling of synapses mimicks the fine 
tuning of wiring that occurs in the developing brain, when neuronal activity 
can modify the synaptic circuitry, once the basic patterns of brain wiring 
are established \cite{alb}.

Our brain is driven by setting the potential of the input site to the value 
$v_{\rm max}$, corresponding to one stimulus. We let the discharge 
evolve until no further firing occurs, then we apply the next stimulus. 
Fig.1 shows the electrical signal as function of time: the total current 
flowing in the system is recorded in time during a sequence of successive 
avalanches. As defined above the time unit corresponds to the time 
necessary to propagate 
the signal from a neuron to next nearest neighbours. Data show that discharges 
of all sizes are present in the brain response, reminiscent of
self-organized criticality where the avalanche size distribution 
scales as a power law \cite{beg}.

\begin{figure}
\includegraphics[width=6cm,angle=270]{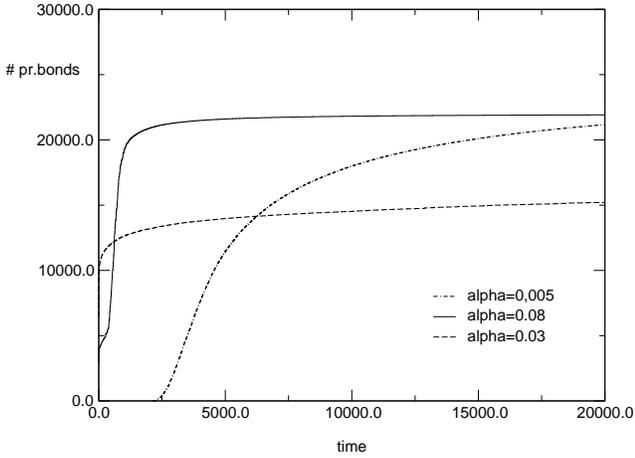}
\caption{ The number of pruned bonds as function of time in a lattice of linear
size $L = 100$
for different values of $\alpha$. The value of the parameters is
$\sigma_t = 0.0001$ and $v_{\rm max} = 6$.
\\}
 \label{fig2}
 \end{figure}

The pruning mechanism introduced in the model depends on the strength of the 
parameter $\alpha$, that controls both the LTP and LTD processes. In fact, 
the more the system learns strengthening the used synapses, the more the 
unused connections will weaken. Fig. 2 shows the number of pruned bonds in 
the system as function of time for different values of $\alpha$: For large 
values of the parameter ( e.g. $\alpha = 0.08$) the system strengthens more 
intensively the synapses carrying current but also very rapidly prunes the 
less used connections, reaching after a short transient a plateau where it 
prunes very few bonds. On the contrary, for small values of $\alpha$ (equal 
to 0.005) the system takes more time to initiate the pruning process and 
slowly reaches a plateau. The number of active (non-pruned) bonds 
asymptotically reaches its largest value at the 
value $\alpha=0.03$. This could be interpreted as an optimal value for the 
system with respect to the joint mechanisms LTP-LTD. Indeed LTD seems to be 
the necessary counterpart to LTP in order to modulate the synaptic strength 
\cite{ros}. 
The asymptotic plateau value for varying $\alpha$ is found within the 
extreme values shown in Fig.2.

After $N_p$ stimuli the network is no longer a simple square lattice 
due to pruning, and constitutes the first approximation to a trained brain, 
on which we are going to perform our measurements. These consist of a new 
sequence of stimuli at the input site, each one of them again triggered by the 
voltage set at threshold, during which we measure the number of firing neurons 
as function of time. This quantity corresponds to the total current flowing in 
a discharge measured by the electromagnetic signal of the EEG. In order to 
compare with medical data, we calculate the power spectrum of the resulting 
time series, i.e. the square of the amplitude of the Fourier transform as 
function of frequency.

In Fig. 3 we show the power spectrum obtained with our model in a log-log-plot 
with the parameters $\alpha= 0.03$, $N_p  = 10$, $\sigma_t = 0.0001$, 
$v_{\rm max} = 6$ and a lattice of size $L = 1000$ and see that it yields a 
power law with the exponent $0.8 \pm 0.1$. This is exactly the same value
for the exponent 
found generically on medical EEG power spectra \cite{fre,nov}. 
We also show in 
Fig. 3 the EEG obtained from channel 17 in the left hemisphere of a male 
subject, as measured in ref.\cite{nov} having the exponent 0.795.
 
\begin{figure}
\includegraphics[width=6cm,angle=270]{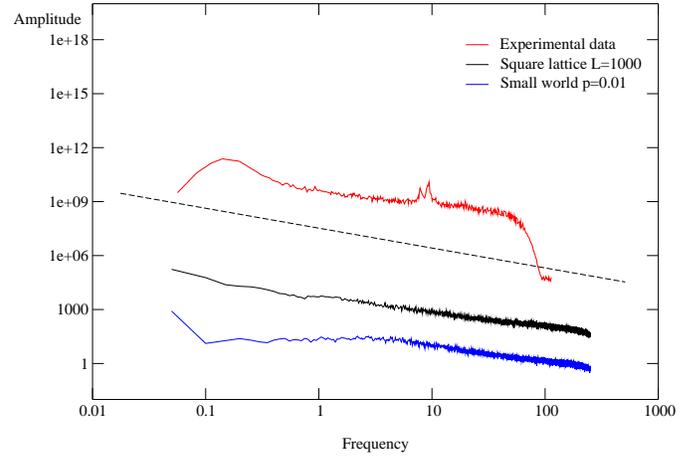}
\caption{Power spectra for experimental data and numerical data ($L = 1000$, 
$\alpha= 0.3$, $N_p = 10$, $v_{\rm max} = 6$) for the square lattice 
(middle curve) and the 
small world lattice (bottom curve, $L = 1000$, $\alpha= 0.05$, $N_p = 1000$, 
$v_{\rm max} = 8$) with $1 \%$ rewired bonds. The experimental data 
(top curve) are from 
ref. (16). The numerical data are averaged over 10 different network 
configurations. The dashed line has a slope 0.8.
\\ }
 \label{fig3}
\end{figure}

The exponent 0.8 that we observed in the power spectrum of our "brain" is 
stable against changes of the parameters $\alpha$, $v_{\rm max}$, $\sigma_t$, 
and $N_p$, and it is also found in the case of random initial conductance 
on the bonds of 
the lattice. We also simulate the brain dynamics on a square lattice with a 
small fraction of bonds, from 0 to $10 \%$, rewired to long range connections 
corresponding to a small world network \cite{wat,lag,she}, which more
realistically reproduces the connections in the real brain. 
Fig.3 shows the power spectrum for a 
system with $1 \%$ rewired bonds and a different set of parameters $\alpha$, 
$N_p$, $v_{\rm max}$: the spectrum has some deviations from the power law at 
small frequencies and tends to the same universal scaling behaviour at larger 
frequencies over two orders of magnitude. The same behaviour is found for a 
larger fraction of rewired bonds, up to $30 \%$. 
 
Although we cannot justify the exponent 0.8 beyond the numerical result of 
our model, it seems clear that this value corresponds to a universal number 
characterizing a larger class of brain networks including real brains. 
Medical studies of EEG focus on subtle details of a power spectrum (e.g. shift 
in peaks) to discern between various pathologies. These detailed structures 
however live on a background power law spectrum that shows universally an 
exponent of about 0.8, as measured for instance in refs. 
\cite{fre} and \cite{nov}. A similar 
exponent was also detected in the spectral analysis of the stride-to-stride 
fluctuations in the normal human gait which can directly be related to 
neurological activity \cite{hau}. We have been able to reproduce this universal 
exponent with a simple electrical toppling model that includes pruning. This 
success is very encouraging since it can provide insight to understand its 
origin and open new perspectives to model pathological features 
of EEG spectra by including more realistic details into our model.

{\small Acknowledgements. We gratefully thank E. Novikov and collaborators for 
allowing us to use their experimental data. We also thank Salvatore Striano, 
MD, for discussions and Stefan Nielsen and Hansj\"org Seybold for help.
}

\end{document}